\documentclass[showpacs,twocolumn,floatfix]{revtex4}
\usepackage{graphicx,psfrag,amsmath,amssymb,amsfonts,latexsym,color,dcolumn,
epsf,graphpap}

\begin{document}

\title{Corrections to Berry's phase in a solid-state qubit due to low frequency noise}

\author{Fernando C. Lombardo\footnote{lombardo@df.uba.ar}
and Paula I. Villar\footnote{paula@df.uba.ar}}
\affiliation{Departamento de F\'\i sica {\it Juan Jos\'e
Giambiagi}, FCEyN UBA and IFIBA CONICET-UBA, Facultad de Ciencias Exactas y Naturales,
Ciudad Universitaria, Pabell\' on I, 1428 Buenos Aires, Argentina}

\date{today}

\begin{abstract}
 We present a quantum open system approach 
to analyze the non-unitary dynamics of a superconducting qubit when it 
evolves under the influence of external noise. 
We consider the presence of longitudinal and transverse environmental fluctuations affecting the system's
dynamics and model these fluctuations by defining their correlation function in time.
By using a Gaussian like noise-correlation, we can study low and high frequency noise 
contribution to decoherence and implement our results in the computation of
geometric phases in open quantum systems. We numerically study
when the accumulated phase of a solid-state qubit can still be found
close to the unitary (Berry) one. Our results can be used to explain experimental measurements of the
Berry phase under high frequency fluctuations and design experimental future setups
when manipulating superconducting qubits.

\end{abstract}

\pacs{05.40.-a;05.40.Ca;03.65.Yz}

\maketitle

\newcommand{\beq}{\begin{equation}}
\newcommand{\eeq}{\end{equation}}
\newcommand{\dalam}{\nabla^2-\partial_t^2}
\newcommand{\mbf}{\mathbf}
\newcommand{\itm}{\mathit}
\newcommand{\beqa}{\begin{eqnarray}}
\newcommand{\eeqa}{\end{eqnarray}}

\section{Introduction}
\label{intro}

Geometric phases are closely linked to the
classical concept of parallel transport of a vector on
a curved surface. This analogy is particularly clear
in the case of a two-level system (a qubit) in the
presence of a biased field that changes in time. Take for example
a spin-$1/2$ particle in a changing
magnetic field. The general Hamiltonian for such
a system is $H= \hbar/2 \vec{R}\cdot \vec{\sigma}$, 
where $\vec{\sigma}=(\sigma_x, \sigma_y,\sigma_z)$
are the Pauli operators and $\vec{R}$ is the biased field vector.
The qubit state can be represented by a point on a sphere of unit
radius, called Bloch sphere. This sphere can be embedded in a three dimensional
space of Cartesian coordinates, and hence the Bloch vector $\vec{R}$ is a vector whose components
$(x,y,z)$ single out a point on the sphere. This representation
offers a particularly well-suited framework to visualize the dynamics of
the qubit, which consists in the qubit state continually precessing about the vector $\vec{R}$,
acquiring a dynamical phase $\gamma(t)$. If the evolution is done adiabatically,
the qubit also acquires a geometric phase (GP), sometimes called Berry phase.

It is known that the system can retain the information of
its motion in the form of this GP, which was first put forward by Pancharatman in
optics \cite{Pancharatman} and later studied explicitly by Berry in
a general quantal system \cite{Berry}. Since then, great progress
has been achieved in this field. The application of the GP has
been proposed in many fields, such as the geometric quantum
computation. Due to its global properties, the GP is
propitious to  construct  fault tolerant  quantum  gates. In this
line of work, many physical systems  have  been  investigated  to
realize  geometric  quantum  computation,  such  as  NMR  (Nuclear
Magnetic Resonance)\cite{NMR}, Josephson  junction \cite{JJ},  Ion  trap \cite{IT} and
semiconductor  quantum  dots \cite{QD}. The quantum computation scheme for the 
GP has been proposed based on the Abelian  or
non-Abelian geometric concepts, and the GP has been shown
to be robust against faults in the presence of some kind of
external noise due to the geometric nature of Berry phase \cite{refs1, refs2, refs3}. 
Then, for isolated quantum systems, the GP is theoretically perfectly understood and 
experimentally verified. However, it was seen that the interactions play an important role for the
realization of some specific operations. As the gates operate slowly compared to the 
dynamical time scale, they become vulnerable to open system effects and parameters' fluctuations
that may lead to a loss of coherence. Consequently, the study of the
GP was soon extended to open quantum systems. Following this idea,
 many authors have analyzed the correction to the GP
 under the influence of an external thermal or non-equilibrium environments, using different
approaches (see \cite{Tong,Gefen, pra, nos, pau, nos13} and references therein). 
In all cases, the purely dephasing model considered was a spin-$1/2$ particle coupled to the
environment's degrees of freedom through a $\sigma_z$ coupling.
The interest on the GP in open systems has also been extended to some experimental setups
\cite{prl}.

The GP is a promising building block for noise-resilient
quantum operations. Lately, the GP has also been observed in a variety of superconducting 
systems \cite{ scienceexp, berger}.
Superconducting circuits are good candidates to potentially manipulate efficiently quantum information. 
 Current circuit technology allows scaling to large and more complex circuits 
\cite{mahn-soo, urbina}.
Several experiments with superconducting Josephson-junction circuits  have demonstrated
quantum coherent oscillations with a long decay time, probing coherent properties of Josephson qubits
and positioning them as useful candidates for applications in quantum computing and 
quantum communication. Despite the long coherence times of the quantum state, the decoherence induced process
still deserves 
study for using these circuits for the development of a quantum processor. 
When the two lowest energy levels of a current-biased Josephson junction are used as a qubit, the qubit state
can be fully manipulated with low and microwave frequency control currents. Circuits presently 
being explored combine in variable ratios the Josephson effect and single Cooper-pair charging effects.
In all cases the Hamiltonian of the system can be written

\begin{equation}
H = \frac{\hbar}{2} \omega_a \sigma_z + \hbar \Omega_{\rm R} \cos(\omega t + \varphi_{\rm R}) \sigma_x,
\label{Hqubit}
\end{equation}
where $\hbar \Omega_{\rm R}$ is the dipole interaction amplitude between the qubit and the microwave 
field of frequency $\omega$ and phase $\varphi_{\rm R}$. $\Omega_{\rm R}/2\pi$ is the Rabi frequency. 
This Hamiltonian can be transformed to a rotating frame at the frequency $\omega$ by means of an
unitary transformation, resulting in a new effective Hamiltonian of the form
\begin{equation}
H_{\text{eff}} = \frac{\hbar}{2} \left(\Delta \sigma_z + \Omega_x \sigma_x + \Omega_y \sigma_y \right) ,
\label{Hqubitdiag}
\end{equation}
where $\Omega_x = \Omega_{\rm R} \cos\varphi_{\rm R}$ and  $\Omega_y = \Omega_{\rm R} \sin\varphi_{\rm R}$. 
This model 
is similar to the generic situation of a qubit in a changing magnetic field,
where ${\bf R} = (\Omega_x,\Omega_y,\Delta)$ 
and $\Delta = \omega_a - \omega$ is 
the detuning between the qubit transition frequency and the applied microwave frequency. 
In an experimental situation \cite{scienceexp} $\Delta$ can be kept fixed 
and one can control the biased field to trace circular paths of different radii $\Omega_{\rm R}$. 

The same physical structures that make these superconducting qubits easy to manipulate, measure, 
and scale are also responsible for coupling the qubit to other electromagnetic degrees of freedom that can 
be a source of decoherence via noise and dissipation. Thus, a detailed mechanism of decoherence and
noise due to the coupling of Josephson devices to external noise sources is still required.
It has been shown that low frequency noise is an important source of decoherence for superconducting qubits.
Generally, this noise is described by fluctuations in the effective magnetic field which are directed
either in the $z$ axis -longitudinal noise- or in a transverse direction -transversal noise.
Both types of noise have been phenomenologically modeled by making different assumptions on these fluctuations,
such as being due to a stationary, Gaussian and Markovian process \cite{berger}. Others, have considered that
the $1/f$ noise must be rooted in a non Gaussian long-time correlation stochastic process. In the context of
quantum information, the implication of long-time correlations of stochastic processes is that the effects
suffered by the system's evolution due to the $1/f$ noise are protocol or measurement dependent.
Apparently, some protocols clearly reveal a non Gaussian nature while others Gaussian approximations attain
the main effects in a short-time scale \cite{paladino}.

In this manuscript,  we shall present  a fully quantum open system approach 
to analyze the non-unitary dynamics of the solid-state qubit when it is considered 
evolving under the influence of external fluctuations. 
We consider the qubit coupled in a longitudinal and transversal directions. As a physical
example, we study the dynamics and decoherence induced process on the superconducting qubit.
We further analyze when  the accumulated phase gained by the system after one period can still be found
close to the unitary (Berry) one and focus on the importance of the longitudinal coupling as a source of
decoherence.
The paper is organized as follows:
in Section II,
 we develop a general quantum open system model in 
order to consider different type of fluctuations (longitudinal and/or transverse) that induce decoherence
on the main system. By means of a general master
equation for the reduced density matrix of the qubit, we follow the non-unitary evolution 
characterized by fluctuations, 
dissipation and decoherence. This gives us a complete insight into the state of the system: complete knowledge
of different  dynamical time-scales and analysis of the effective role of noise sources
inducing decoherence. 
Section III contains the 
numerical evaluation of the geometric phase and its 
noise induced corrections for the several scenarios considered. 
We shall emphasize the effect of longitudinal and transversal noise on the global 
geometric phase. Comparison between theory and experiment verifies our understanding of the physics 
underlying the system as a 
dissipative two-level device.  Berry's phase measurements provide an important constraint to take into account about noise models 
and their correction induced 
over the GP, at least, at the times in which the experiments can be performed. The comprehension of the
decoherence and dissipative processes should allow their further suppression in future qubits designs or 
experimental setups.
In Section IV we summarize our findings.


\section{Master equation approach to decoherence in a superconducting qubit}
\label{Deco}

We shall begin by deriving a general master equation for the reduced density matrix
for the qubit (obtained after tracing out all the 
environmental degrees of freedom).
The dynamics of a generic two-level system steered by a system's Hamiltonian of the 
type (where we have set $\hbar = 1$ all along the paper)

\begin{eqnarray}
H_{\text{Total}} &=& H_q + H_{\rm int} + H_{\cal E}, ~~\text{with}\\
H_q&=& \frac{1}{2} \left( \Omega \sigma_x + \Delta \sigma_z \right) \label{Hq}
\end{eqnarray}
where we have defined a qubit Hamiltonian $H_q$ similar to that of a solid-state qubit
Eq.(\ref{Hqubitdiag}) - setting 
$\varphi_{\rm R} = 0$ for simplicity-, and $H_{\cal E}$ is the Hamiltonian of the bath.
The interaction
Hamiltonian is thought as some longitudinal and transverse noise coupled to the main system:

\begin{equation}
H_{\rm int} = \frac{1}{2} \left( {\hat {\delta\omega_1}} \sigma_x + {\hat {\delta\omega_0}} \sigma_z\right).
\label{Hint}
\end{equation}

We must note that the system's unitary dynamics and coupling to the environment is different from
the usual purely dephasing models proposed to study geometric phases in open systems Refs.
\cite{Tong,Gefen, pra, nos, pau, nos13,prl}.
We shall derive the master equation in the Born-Markov approximation, for general noise terms 
${\hat {\delta\omega_1}}$ and ${\hat {\delta\omega_0}}$ interacting with the system in
the $\hat{x}$ and $\hat{z}$ directions, respectively. We will consider a weak coupling 
between system and environment and that the bath is sufficiently large to stay in a  
stationary state. In other words, the total state $\rho_{\cal SE}$ (system and environment) 
can be split as 

\begin{equation}
\rho_{\cal SE} \approx \rho(t) \times \rho_{\cal E}  ,
\end{equation}
for all times. It is important to stress that due to the Markov regime, we will restrict to cases for which 
the self-correlation functions generated at the environment (due to the coupling interaction) would 
decay faster than typical variation scales in the system. In this way, the evolution equation 
for $\rho(t)$ is local 
in time \cite{breuer}. In the interaction picture, the evolution of the total state 
is ruled by the Liouville equation

\begin{equation}
{\dot \rho}_{\cal SE} = -i \left[ H_{\rm int}, \rho_{\cal SE} \right],
\end{equation} where we have denoted the state $\rho_{\cal SE} $ in the interaction picture in the same 
way than before, just in order to simplify notation. 
A formal solution of the Liouville equation can be obtained perturbatively using the Dyson expansion \cite{jpp}:

\begin{eqnarray}
\rho_{\cal SE} (t) &=& \sum_{n\ge 0} \int_0^t ds_1\int_0^{s_1} ds_2 .... \int_0^{s_n} ds_n 
\left(\frac{1}{i}\right)  \\
&\times& \left[H_{\rm int}(s_1),\left[ H_{\rm int}(s_2),\left[ .... , \left[H_{\rm int}(s_n),
\rho_{\cal SE} (0)\right] ... \right]\right]\right]. \nonumber
\end{eqnarray}

From this expansion, one can obtain a perturbative master equation, up to second order in 
the coupling constant between system and 
environment for the reduced density matrix $\rho = {\mbox Tr}_{\cal E} \rho_{\cal SE}$. In the 
interaction picture the formal solution reads as

\begin{eqnarray}
\rho (t) &\approx & \rho (0) - i \int_0^t ds {\mbox Tr}_{\cal E} \left(\left[H_{\rm int}(s), 
\rho_{\cal SE}  (0)\right]\right) \\
&-&\int_0^{t}ds_1 \int_0^{s_1} ds_2  {\mbox Tr}_{\cal E} \left(\left[H_{\rm int}(s), 
\left[H_{\rm int}(t), \rho_{\cal SE} (0)\right]\right]\right).
\nonumber \end{eqnarray}

Taking the temporal derivative of the previous equation, and assuming that system and bath are 
not correlated at the initial time, the master equation can be written as \cite{breuer}

\begin{eqnarray}
&&\dot\rho = -i \, {\mbox Tr}_{\cal E}\left[H_{\rm int}(t),\rho(t)\times\rho_{\cal E}(0)\right] 
\nonumber \\
&-& \int_0^t ds \, {\mbox Tr}_{\cal E}\left[H_{\rm int}(t),\left[H_{\rm int}(s),\rho(t)\times\rho_{\cal E}(0)
\right]\right] \nonumber \\
&+& \int_0^t ds {\mbox Tr}_{\cal E}\left( \left[H_{\rm int}(t),{\mbox Tr}_{\cal E}\left(\left[H_{\rm int}(s),
\rho(t)\times\rho_{\cal E}(0)\right]\right) \right.\right. \nonumber \\
 && \left.\left.  \times  \rho_{\cal E}(0)\right]\right) .\nonumber \end{eqnarray}

Considering that the ${\hat {\delta\omega_i}}$ of the $H_{\rm int}$ (Eq.(\ref{Hint}))
are operators acting only on the Hilbert space of the environment
(and the Pauli matrices applied on the system Hilbert space), the master equation, in the Schr\"odinger picture,
can be written as
\begin{equation}
\dot \rho = - \int_0^t ds \, {\mbox Tr}_{\cal E} \left[ H_{\rm int} (t) , \left[ H_{\rm int} (s), 
\rho (t) \times \rho_{\cal E}(0)\right] \right]. 
\end{equation}

The master equation explicitly reads

\begin{eqnarray}
\dot\rho &=& -i \left[H_q, \rho\right] - D_{xx}(t) \left[\sigma_x,\left[\sigma_x,\rho\right]\right] 
- f_{xy}(t) \left[\sigma_x,\left[\sigma_y,\rho\right]\right] \nonumber \\ &-& f_{xz}(t) 
\left[\sigma_x,\left[\sigma_z,\rho\right]\right] 
-  f_{zx}(t) \left[\sigma_z,\left[\sigma_x,\rho\right]\right] \nonumber \\ &-& f_{zy}(t) 
\left[\sigma_z,\left[\sigma_y,\rho\right]\right] - D_{zz}(t) \left[\sigma_z,\left[\sigma_z,\rho\right]\right], 
\label{mastereq}
\end{eqnarray}
where the noise coefficients are given by

\begin{eqnarray}
D_{xx}(t) &=& \int_0^t ds \, \langle {\hat {\delta\omega_1}}(0) {\hat {\delta\omega_1}}(-s)\rangle_{\cal E}  \, X_1(-s) \nonumber \\
f_{xy}(t) &=& \int_0^t ds \, \langle {\hat {\delta\omega_1}}(0) {\hat {\delta\omega_1}}(-s)\rangle_{\cal E}  \, Y_1(-s) \nonumber \\
f_{xz}(t) &=& \int_0^t ds \, \langle {\hat {\delta\omega_1}}(0) {\hat {\delta\omega_1}}(-s)\rangle_{\cal E}  \, Z_1(-s) \nonumber \\
f_{zx}(t) &=& \int_0^t ds \, \langle {\hat {\delta\omega_0}}(0) {\hat {\delta\omega_0}}(-s)\rangle_{\cal E}  \, X_0(-s) \nonumber \\
f_{zy}(t) &=& \int_0^t ds \, \langle {\hat {\delta\omega_0}}(0) {\hat {\delta\omega_0}}(-s)\rangle_{\cal E}  \, Y_0(-s) \nonumber \\
D_{zz}(t) &=& \int_0^t ds \, \langle {\hat {\delta\omega_0}}(0) {\hat {\delta\omega_0}}(-s)\rangle_{\cal E}  \, Z_0(-s). \label{coeficientes}
\end{eqnarray}
It is possible to recognize $D_{ab}$ and $f_{ab}$ as normal and anomalous diffusion coefficients,
respectively ($a,b=x,y,z$). The 
functions $X_{0,1}, Y_{0,1}$, and $Z_{0,1}$ are derived by obtaining the temporal dependence of the 
Pauli operators $\sigma_i$ in the Heisenberg representing through the differential equations,
\begin{equation}\frac{d\sigma_k(t)}{dt} = i \left[H_q, \sigma_k(t)\right],
\end{equation}
with $k = x, y, z$ and $H_q$ as in Eq.(\ref{Hq}). The solution can be expressed as a 
linear combination of the Pauli matrices  (in the Schr\"odinger
representation), 
\begin{equation} \sigma_z^{0,1} = X_{0,1}(t) \sigma_x + Y_{0,1}(t) \sigma_y  + Z_{0,1}(t) \sigma_z.
\end{equation} 
The solution can be easily written as
\begin{eqnarray}
X_1(t) &=& \frac{\Omega^2 + \Delta^2 \cos(2 t \sqrt{\Omega^2 + \Delta^2})}{\Omega^2 + \Delta^2},\nonumber \\
Y_1(t) &=& \frac{\Delta \sin(2 t \sqrt{\Omega^2 + \Delta^2})}{\sqrt{\Omega^2 + \Delta^2}},\nonumber \\
Z_1(t) &=& X_0(t) = \frac{\Delta \Omega \left[ 1 - \cos(2 t \sqrt{\Omega^2 + \Delta^2})\right] }{\Omega^2 + \Delta^2},\nonumber \\
Y_0(t) &=& - \frac{\Omega \sin(2 t \sqrt{\Omega^2 + \Delta^2})}{\sqrt{\Omega^2 + \Delta^2}},\nonumber \\
Z_0(t) &=& 1 - \frac{\Omega^2 \left[ 1 - \cos(2 t \sqrt{\Omega^2 + \Delta^2})\right] }{\Omega^2 + \Delta^2}
.\nonumber 
\end{eqnarray}
It is easy to check that if the Rabi frequency is zero and $\delta \hat \omega_1 =0$, 
we recover the dynamics of a spin-$1/2$ precessing a bias field vector ${\bf R}$.

The idea is to use different noise correlation functions  to model
different types of noise that can be found in solid-state qubits. Once the coefficients in 
Eqs.(\ref{coeficientes})
are defined, we can numerically solve the master equation and obtain the evolution 
in time of the reduced density matrix. 
Once this quantity is known, we can further obtain interesting features of the qubit dynamics
such as the biased vector ${\bf R}$ and the decoherence induced on the superconducting qubit.

The noise correlations can be defined by their spectral density $J_i(\omega) = 1/(2\pi)
\int dt e^{i \omega t} 
\langle {\hat {\delta\omega_i}}(0) {\hat {\delta\omega_i}}(-s)\rangle_{\cal E}$ with 
$i = 0, 1$. 
 Herein, we shall focus on the long and short-correlated noise 
(slow and sharp decay of 
$\langle {\hat {\delta\omega_i}}(0) {\hat {\delta\omega_i}}(-s)\rangle$), i.e. on the noise 
power peaked at low or high frequencies. We will describe different types of noise
as 
\begin{equation}
\langle
{\hat {\delta\omega_i}}(0) {\hat {\delta\omega_i}}(-s)\rangle_{\cal E} = \gamma_i {\cal F}(\alpha_i, t)
\label{F}
\end{equation}
(where $\gamma_i$ is a dissipative constant that includes the coupling strength
between system and bath, 
and $\alpha_i$ is a parameter with frequency units). This function ${\cal F}$ keeps the information about 
the correlation times and couplings in the environment.  Phenomenologically, ${\cal F}$  can be
thought as a Dirac delta functional for short-correlations in time-domain, or
a Gaussian-like function of time for a more general scenario. 
In solid-state systems decoherence is potentially strong due to numerous microscopic modes. Noise is 
dominated by material-dependent sources, such as background-charge fluctuations or variations of magnetic 
fields and critical currents, with given power spectrum, often known as $1/f$. This noise is 
difficult to suppress and, since the dephasing is generally dominated by the low-frequency noise, it is
particularly destructive (though it is said that can be reduced by tuning the linear longitudinal 
qubit-noise coupling 
to zero \cite{vion}). A further relevant contribution is 
the electromagnetic noise of the control circuit, typically Ohmic at low frequencies.

{\it Gaussian noise}. An interesting way to model the fluctuations is through a Gaussian-correlated noise.
We assume that the operator ${\hat {\delta\omega_i}}(t)$, 
is given by a random function 
$\delta\omega_i (t)$ with $\langle {\hat {\delta\omega_i}}(t)\rangle_{\cal E} = 0$ 
and its correlation between the values of 
$\delta\omega_i (t)$ at two different times is non-zero only for this time interval. Explicitly, 

\begin{figure}[!ht]
\includegraphics[width=8.cm]{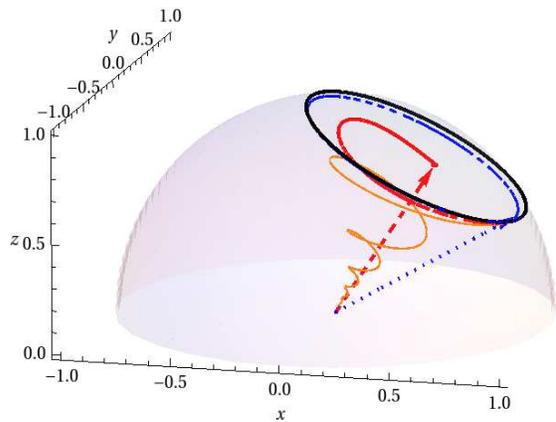}
\caption{(Color online) The evolution of the system can be illustrated by the path 
traversed by the vector ${\bf R}$ in the Bloch sphere. Solving the master equation 
it is possible to analyze the decoherence process by mean of the change in the absolute
value of ${\bf R}$, which implies the 
loss of purity of the system, and also its change of ${\hat z}$-component. The black 
dotted line corresponds to the unitary evolution, i.e. when the qubit evolves isolated 
from the environment in a circle on the sphere surface. The orange trajectory, which approaches 
the center of the sphere, corresponds to the $\delta$-correlated noise with 
$\gamma_0=\gamma_1=0.03 \Delta$. We can see that after a few number of periods,
system looses coherence completely and the final state, is a totally mixed one.  
Red and blue trajectories correspond to different values of parameters $\alpha_0$ 
and $\alpha_1$ of the Gaussian-correlated noise models. Red curve shows a more 
decoherent behavior, and corresponds to a low value of $\alpha_0=\alpha_1=0.03 \Delta$. 
Blue line, to higher values $\alpha_0=\alpha_1=30\Delta$. We can see that the slow 
decaying of noise correlations, the more decoherece on the qubit in the weak coupling 
case $\gamma_0=\gamma_1=0.03 \Delta$. We have set $\Omega = 0.5 \Delta$. }
\label{Fig1}
\end{figure}

\begin{equation}
\langle {\hat {\delta\omega_i}}(t_1) {\hat {\delta\omega_i}}(t_2)\rangle_{\cal E} = \Phi_i(t_1 - t_2),
\end{equation}
where $\Phi_i(t)$ is a function sharply peaked at $t=0$ and vanishing for $t > \tau_{\rm c}$ for 
a critical time-scale $\tau_{\rm c}$.  
We have set  $\Phi_i(t) = \gamma_i {\cal F}(\alpha_i, t)$ as defined in Eq.(\ref{F}) where
${\cal F}$ is a Gaussian-like function of time. By setting the parameter  $\alpha_i$ ($\alpha_0$ for the longitudinal
noise since it affects the coupling in $\hat z$ axis and $\alpha_1$ the transverse noise -coupling in $\hat x$-)
of the model, we can study low or high frequency noise contribution to decoherence. 
Therefore, in this case, decoherence depends on the interplay of $\alpha_0$ and $\alpha_1$ 
and the value of the dissipation constants $\gamma_0$ and $\gamma_1$.  
For example, in Fig.\ref{Fig1} we present the trajectory of the Bloch vector during a cyclic (or quasicyclic) evolution.
The black circle on the surface of the Bloch sphere is the evolution of the vector $\bf{R}$ in the
unitary case, i.e. $\gamma_0=0=\gamma_1$. Herein, we see that in absence of environment the qubit performs 
a closed trajectory in a period $\tau$, acquiring the known GP, $\phi_G= \pi(1-\cos(\vartheta))$ with 
$\vartheta=\Delta/(\sqrt{\Delta^2 + \Omega^2})$.  
By considering different values for the parameters of our noise model: $\gamma_i$ and $\alpha_i$, we
can evaluate how the distinct environments affect the system's dynamics.
In Fig.\ref{Fig1}, we also present the different trajectories of the
Bloch vector $\bf{R}$ for a value of $\gamma_0=0.03\Delta$ and $\gamma_1=0.03\Delta$. As $\gamma_i$ are related to the
square of the coupling constant, these values for $\gamma_i$ represent a significant environment within the weak coupling 
approximation. The blue dotted line is the trajectory of the Bloch vector when $\alpha_0=30\Delta$ and 
$\alpha_1=30\Delta$. This trajectory is very similar to the unitary one, which means that the environment has little 
influence on the systems' dynamics. The blue arrow line that starts in the center of the sphere and goes to the surface indicates
the position of the Bloch sphere after one cycle $\tau=2\pi/\tilde{\Omega}$, $\tilde{\Omega}= \Delta/
\sqrt{\Delta^2 + \Omega^2}$. The red solid arrow line is the trajectory  for a low value of $\alpha_0=0.03\Delta=\alpha_1$.
This is what we shall call low frequency noise. In this case, we can note that the trajectory differs substantially
from the unitary one, meaning the system's dynamics is affected by the decoherence process. 
 Qualitatively, decoherence can be thought of as the deviation of probabilities 
measurements from the ideal intended outcome. Therefore, 
decoherence can be understood as fluctuations in the Bloch vector ${\bf R}$ induced by noise. 
Since decoherence rate depends on the state of 
 the qubit, we will represent decoherence by the change of $\vert{\bf R}\vert$ in time, 
starting from $\vert {\bf R}\vert = 1$ for the initial 
 pure state, and decreasing as long as the quantum state losses purity. 
The red dashed 
 Bloch vector after a cycle is not longer on the surface of the sphere as can be seen in Fig.\ref{Fig1}.
The module of the red dashed Bloch vector
has been reduced $16 \%$ after one cycle with respect to the module of the unitary Bloch vector.

As a particular case, we can mention a noise correlation function given by a 
function ${\cal F}= \delta(s)$. If the general environment considered in this approach is a bath of
harmonic oscillators with a delta-correlation function ($J(\omega) \sim \omega$), 
then we will be modeling an
ohmic bath in the limit of finite temperature \cite{pra}. This assumption implies that
the only  coefficients in Eq. (\ref{mastereq}) which are constant and non zero
are $D_{zz}=\gamma_0 k_BT$ and $D_{xx}=\gamma_1k_B T$. This model is commonly known as dephasing. This modeling of the 
environment is also included in Fig.\ref{Fig1} for $\gamma_0=\gamma_1=0.03 \Delta$ with an orange line.
It is easy to see that the Bloch vectors decays to the center of the sphere loosing purity faster 
than in the Gaussian model. In the latter, due to the presence of more terms in the master equation,
the Bloch vector does not decay to the center of the sphere \cite{benenti}.

\begin{figure}[!ht]
\includegraphics[width=8.cm]{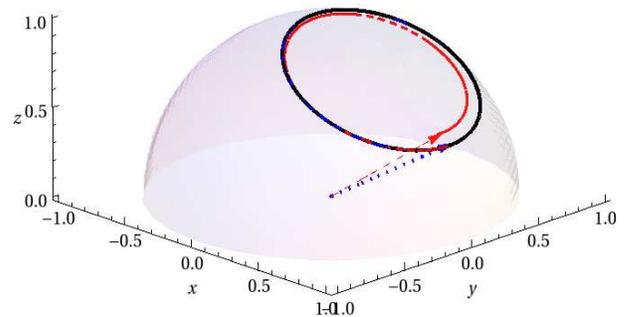}
\caption{(Color online) Numerical solution of the master equation for the trajectory 
of vector  ${\bf R}$ in the Bloch sphere,  for the Gaussian noise models for smaller 
dissipative constants $\gamma_0=\gamma_1=0.03 \Delta$. As before,  black solid 
line corresponds to the unitary evolution. Red and Dotted blue trajectories correspond 
to different values of parameters $\alpha_0$ and $\alpha_1$ of the Gaussian-correlated 
noise models. Red curve shows a more decoherent behavior due to a low 
value of $\alpha_0=\alpha_1=0.03 \Delta$. Blue line corresponds to higher values 
$\alpha_0=\alpha_1=30\Delta$. We have set $\Omega = 0.5 \Delta$. }
\label{Fig2}
\end{figure}

In Fig.\ref{Fig2}, we present a different scenario since the trajectories presented correspond to a very weak
environment $\gamma_0=\gamma_1=0.03 \Delta$. Once again, the black solid line is the reference for the unitary
case while the blue line (almost coincident with the black) is for high frequency ($\alpha_0=\alpha_1=30\Delta$)
and the red one for low frequency noise ($\alpha_0=\alpha_1=0.003 \Delta$). Here, the Bloch vector for the low frequency
noise (red) is $5\%$ reduced with respect to the unitary Bloch vector after one cycle $\tau$.

\begin{figure}[!ht]
\includegraphics[width=8.cm]{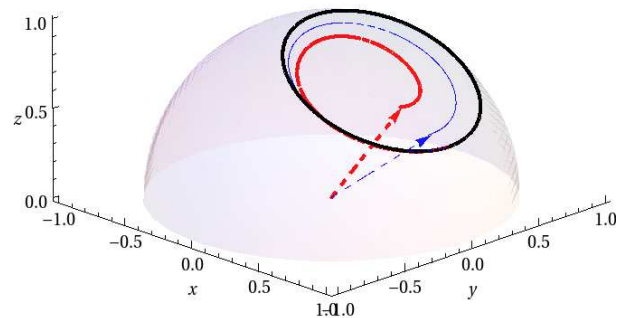}
\caption{(Color online) Numerical solution of the master equation for the trajectory of vector  
${\bf R}$ in the Bloch sphere for the $1/f$ noise model. The black solid line indicates the
trajectory of the qubit in absence of environment. The blue dotted line is the trajectory of the qubit
under the influence of an environment with a ``high'' infrared cutoff $\Lambda=0.1 \Delta$. The red curve
is the trajectory for a ``low'' infrared cutoff $\Lambda=0.001 \Delta$. We have set $\Omega = 0.5 \Delta$.}
\label{Fig3f}
\end{figure}

Finally we can comment the $1/f$ noise mentioned above. This 
noise can be modeled by a bath composed of an infinite set of harmonic
oscillators (similarly to what has been done in the spin boson model \cite{pra}). 
At $T=0$, the noise kernel $\nu(t)$ can be evaluated when $J(\omega) \sim A/\omega$. 
Then, the $1/f$ noise is determined by a correlation function  $\nu(t) = -\gamma \Lambda\,{\mbox CI}(\Lambda t)$,
where ${\mbox CI}(x)$ is the cosine integral function, and $\Lambda$ is the typical infrared cutoff for the
$1/f$ noise. In the high temperature limit, this kernel is given by $\nu(t) = T \, \gamma \Lambda\,
(-\pi/2 \,t + \cos(\Lambda t)/\Lambda + t \, {\mbox SI}(\lambda t))$, with ${\mbox SI}(x)$ 
the sine integral function. This quantitative modeling of the $1/f$ noise through a master equation
approach is somewhat analogous to the effect of the phenomenological modeling of the 
noise through an ensemble of ``spin-fluctuators'' \cite{paladino}. In Fig.\ref{Fig3f} we effectively
note how harmful this type of noise is for the dynamics of the qubit, even in the very low temperature limit. Therein, the black solid line
represents the unitary trajectory of the Bloch vector. In this model, the relevant parameter is the 
infrared frequency cutoff $\Lambda$. The blue dotted line is for a big value of the infrared cutoff 
$\Lambda=0.1 \Delta$,
while the red solid line is for a low frequency cutoff $\Lambda=0.001 \Delta$. Both cases are affected by
decoherence. In the low frequency cutoff case the module of the Bloch vector -indicated as a dashed red arrow from the center 
of the sphere- is reduced  $20\%$ in a cycle $\tau$.


\section{Application: geometric phase of a solid-state qubit in a non-unitary evolution} 

Practical implementations of
quantum computing are always done in the presence of decoherence.
Thus, a proper generalization for the geometric phase to nonunitary 
evolutions is central in the
evaluation of the robustness of geometric quantum computation.
This generalization has been done in
\cite{Tong}, where a functional representation of GP was proposed, 
after removing the dynamical phase from the total phase
acquired by the system under a gauge transformation.

The GP for a mixed state under nonunitary evolution 
is then defined as 
\begin{eqnarray} \Phi &=&
{\rm arg}\{\sum_k \sqrt{ \varepsilon_k (0) \varepsilon_k (\tau)}
\langle\Psi_k(0)|\Psi_k(\tau)\rangle \nonumber \\
&\times & e^{-\int_0^{\tau} dt \langle\Psi_k|
\frac{\partial}{\partial t}| {\Psi_k}\rangle}\}, \label{fasegeo}
\end{eqnarray}
where $\varepsilon_k(t)$ are the eigenvalues and
 $|\Psi_k\rangle$ the eigenstates of the reduced density matrix
$\rho$, solution of the master equation. In the last definition, $\tau$ denotes a time 
after the total system completes
a cyclic evolution when it is isolated from the environment.
Taking the effect of the environment into account, the system no
longer undergoes a cyclic evolution. However, we will consider a
quasicyclic path ${\cal P}:t ~\epsilon~[0,\tau]$ with
$\tau=2 \pi/\tilde{\Omega}$ \cite{Tong}.
 It is worth
noting that the phase in Eq.(\ref{fasegeo}) is manifestly gauge
invariant, since it only depends on the path in the state space,
and that this expression, even though is defined for non
degenerate mixed states, corresponds to the unitary geometric
phase in the case that the state is pure (closed system).

It is expected that Berry's phase can be only observed in experiments
carried out in a time scale slow enough to ignore nonadiabatic corrections, 
 but rapid enough to avoid destructive decoherence \cite{nos}. 
 The noise induced corrections to the GP depend on the value of parameters
present in the noise model, for example $\alpha_i$ and $\gamma_i$ used in the above section. 
 The purpose 
of this section is twofold: study how the GPs are affected by the different models of noise
and explain some recent experimental setups where the GP has been measured in presence 
of noise \cite{scienceexp,berger}. In the mentioned works, authors observed the
Berry's phase in a superconducting qubit by different approaches. However, both experiments
agree on the fact that the longitudinal noise affects the system's dynamics in a clearer way that
the transversal noise. Another important fact is that in \cite{scienceexp} authors claimed
to have observed the Berry phase under high-frequency fluctuations. They considered that this
robustness of the GPs to high-frequency noise may be exploitable in the realization of logic
quantum gates for quantum computation. Therefore, we aim to explain these features of the GP
for our model from a primary derivation of a master equation approach.
In the following, we shall use the Gaussian model of noise for the study of the GP since it is
widely said that the $1/f$ can be reduced in spin-echo experiments, by tuning the linear longitudinal qubit-noise 
to zero \cite{vion}.  In our gaussian model, we have shown that the decoherence process was very
dependent on the value of the $\alpha_i$ parameter, which we associated to a frequency. In all cases shown,
decoherence was enhanced in the low frequency case (small values of $\alpha$,
see Figs. \ref{Fig1} and \ref{Fig2}).

\begin{figure}[!ht]
\includegraphics[width=8cm]{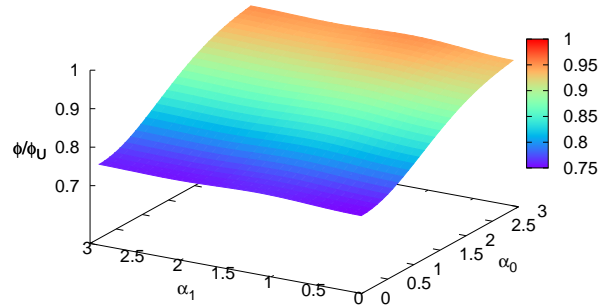}
\caption{(Color online) Ratio between the computed GP in presence of noise and the one computed in the isolated
case $\Phi_U$, as function of $\alpha_0$ and $\alpha_1$ (in units 
of $\Delta$), with $\gamma_0=\gamma_1 = 0.001 \Delta$. The GP is more affected
by the presence of longitudinal noise frequency $\alpha_0$ since the rate is bigger. 
The GP does not considerably depend 
on the transversal noise $\alpha_1$. We have set $\Omega = 0.5 \Delta$. }
\label{Fig5}
\end{figure}

In Fig.\ref{Fig5} we present the ratio between the GP $\Phi$
computed for a system evolving under a noisy environment after a cycle $\tau$
and the unitary one $\Phi_U$, for  different values of $\alpha_i$, having $\gamma_i$ fixed
as $\gamma_0=\gamma_1 = 0.001 \Delta$. We show how this ratio varies once you have a fixed environment
and a tunable frequency. Herein, we can note that the ratio does not practically change for different
values of $\alpha_1$, meaning that the transversal fluctuations are not relevant. However, we can see that
the ratio varies considerably in the $\alpha_0$ direction. The GP is visibly corrected for small values of $\alpha_0$, i.e.
for low frequency noise in the longitudinal coupling of the qubit. This correction means that the Bloch vector
has a relevant difference with the initial unitary Bloch vector since the environment induces more decoherence
in the low frequency case (see Fig.\ref{Fig1}).

\begin{figure}[!ht]
\includegraphics[width=8cm]{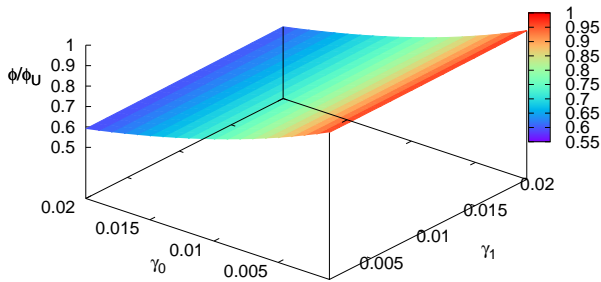}
\caption{(Color online) Rate between the computed GP $\Phi$ in presence of noise and the one computed in the isolated
case $\Phi_U$ as function of the dissipative constants $\gamma_0$
and $\gamma_1$ (in units of $\Delta$), for a fixed value of  $\alpha_0 = \alpha_1 = 0.01 
\Delta $. The ratio is more affected by the longitudinal noise. We have set $\Omega = 0.5 \Delta$. }
\label{Fig6}
\end{figure}

In Fig.\ref{Fig6} we again present the ratio between the GP $\Phi$
computed for a system evolving under a noisy environment after a cycle $\tau$
and the one unitary computed $\Phi_U$. This time we show how this ratio varies for different 
values of $\gamma_0$ and $\gamma_1$ for small values of $\alpha_i$, say $\alpha_0=\alpha_1= 0.01 
\Delta$. It is easy to note that the GP $\Phi$ is very similar to the unitary GP $\Phi_U$, in absence
of longitudinal noise ($\gamma_0=0$), which means that the evolution is not considerably affected
by the transverse noise. However, we can see a different behavior if we consider longitudinal noise ($\gamma_1=0$).
The GP varies perceptibly as the environment is coupled in the longitudinal direction is stronger (bigger values 
of $\gamma_0$). It is important to say that the relevant role of the tunable frequency $\alpha_i$ makes sense
if we are dealing with a considerable environment which can effectively induce noise into our system's dynamics. For 
very small values of $\gamma_0$, Fig.\ref{Fig6} shows that the GP computed is similar to the unitary GP, independently
of low or high frequency fluctuations.

\begin{figure}[!ht]
\includegraphics[width=8cm]{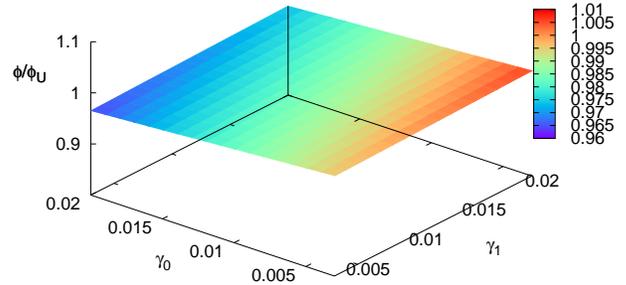}
\caption{(Color online) Rate between the computed GP in presence of noise and the one computed in the isolated
case $\Phi_U$ as function of the dissipative constants $\gamma_0$ and 
$\gamma_1$ (in units of $\Delta$), for a fixed value of  $\alpha_0 = \alpha_1 = 10 \Delta$. 
The correction to the GP is almost negligible for higher values of $\alpha_i$,
for weak coupling with the environment. We have set $\Omega = 0.5 \Delta$. }
\label{Fig7}
\end{figure}

In Fig.\ref{Fig7} we present the ratio between the GP $\Phi$
computed for a system evolving under a noisy environment after a cycle $\tau$
and the one unitary computed $\Phi_U$ as a function of  $\gamma_0$ and $\gamma_1$ for bigger values of 
$\alpha_i$, say $\alpha_0=\alpha_1= 10\Delta$. Herein, we see that the system evolution in
the presence of an environment with high frequency fluctuations is very similar to the unitary evolution, since
the GP acquired is practically similar to the $\Phi_U$, for almost all values of $\gamma_0$. If we get a closer look, we
can note that the difference between both phases becomes slowly to increase for stronger values of $\gamma_0$.
We believe that the situation depicted in Fig.\ref{Fig7} is very similar to the experimental situation reported
in \cite{scienceexp} where authors have measured the Berry phase for a superconductiong qubit under 
high frequency fluctuations.

\begin{figure}[!ht]
\includegraphics[width=8.5cm]{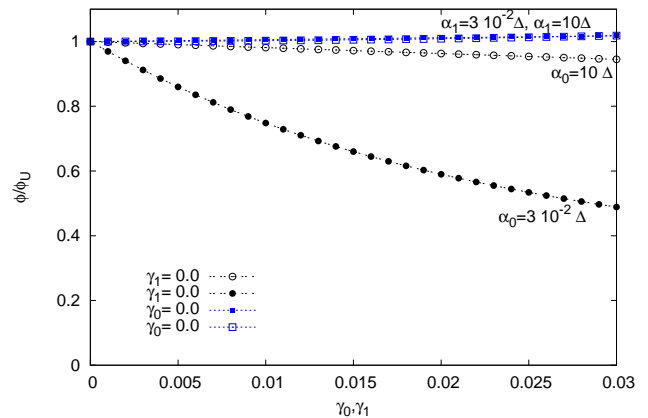}
\caption{(Color online) Ratio of the computed GP $\Phi$ in the presence of a noisy
environment and the unitary GP $\Phi_U$ as a function 
of $\gamma_0$ and $\gamma_1$.  Blue  Square-dot line  is the correction as a 
function of the transverse noise $\gamma_1$, for $\gamma_0 = 0$. Black circle-dotted line 
is the correction to the GP as a function of the 
longitudinal noise $\gamma_0$, when $\gamma_1 = 0$. Noise in the ${\hat z}$-direction corrects the phase 
more than noise in the transversal directions. 
These corrections are in agreement with the behavior 
of decoherence as a function of dissipative constants. All $\gamma_i$ are measured in units of $\Delta$.}
\label{Fig8}
\end{figure}
Finally, in Fig.\ref{Fig8} we quantitatively show how the GP is affected by the longitudinal and transverse
noises separately. We present the ratio between the observed GP $\Phi$ after a cycle $\tau$  
and the unitary GP $\Phi_U$
 as a function of both dissipative constants, $\gamma_i$. We consider that the qubit is coupled to only one
 noise, i.e. that when we show how the ratio varies as function of $\gamma_0$, the qubit is evolving only under
 a longitudinal noise and $\gamma_1=0$ (black circled-line). If the ratio varies as a function of $\gamma_1$, 
 then the qubit is suffering
 the presence of transversal fluctuations only $\gamma_0=0$ (blue squared-lines). 
  We have also add the $\alpha_i$ parameter to have the full scenario.
The correction to the GP is almost 
 imperceptible to  low and high frequency transversal fluctuations (full and empty squares with 
 $\alpha_1 = 0.03\Delta$ and $\alpha_1 =  10 \Delta$ respectively), 
 at least in the weak coupling limit. On the contrary, if the fluctuations of the environment are
 longitudinal, only those high-frequency ones does not considerably affect the measurement of the geometric
 phase. It is evident that low frequency longitudinal noise
 induces a bigger correction to the phase (as can be seen from the full black-circled line with 
 $\alpha_0 = 0.03\Delta$). 
 These results agree with the previous analysis done on decoherence induced in
the qubit and with the experimental
setups reported of the observed geometric phase \cite{scienceexp, berger}.  
It is important to emphasize that our approach is general and allows several ways of modeling the
environment coupled to the main system.

\section{Final Remarks}
\label{conclu}
We have considered the effective two-state Hamiltonian  for the current-biased Josephson junction. 
The qubit 
has been shown to be fully manipulated with the control currents. Like any other quantum object, the 
qubit is subject to decoherence due to the interaction with uncontrolled degrees of freedom in its 
environment, including those in the device itself. These degrees of freedom 
appear as noise induced in the parameters entering the qubit Hamiltonian and also as 
noise in the control currents. These noise sources produce decoherence 
in the qubit, with noise, mainly, at microwave frequencies affecting the relative population between 
the ground and excited state, 
and noise or low-frequency fluctuations affecting the phase of the qubit. It is important to study the 
physical origins of decoherence by 
means off noise spectral densities and noise statistics. 

We have derived a master equation for the two-level system including the combined effect of noise in
the longitudinal and 
transversal directions. We considered different types of noise by defining their correlation function in time.
We have mainly analyzed a Gaussian-like correlated type of noise, with 
low and fast decaying times that induce different decoherence processes in the low or 
high frequency parts of the environmental spectrum. We have even presented very correlated noise, where 
the noise kernel is proportional to a Dirac delta function in time and the $1/f$ known commonly used in 
spin fluctuators environments. For each type of noise presented, we numerically solved the master equation
and obtained the system's dynamics. Qualitatively, decoherence can be thought of as the deviation of probabilities 
measurements from the ideal intended outcome. Therefore, 
decoherence can be understood as fluctuations in the Bloch vector ${\bf R}$ induced by noise. 
Since decoherence rate depends on the state of 
 the qubit, we have represented decoherence by the change of $\vert{\bf R}\vert$ in time, 
starting from $\vert {\bf R}\vert = 1$ for the initial 
 pure state, and decreasing as long as the quantum state losses purity. 

We have extended our analysis of decoherence to understand the corrections induced in the geometric phase, when
the qubit evolves in time under fluctuations of the environment. Within the general picture of
the master equation, we provide a framework to understand when the accumulated 
phase can still be found close to the unitary 
(Berry) one. We have focused on the effect of longitudinal and transversal noise on the global geometric phase. 
It is important to note that the relevant role of the tunable frequency $\alpha_i$ in our gaussian model 
makes sense if we are dealing with a considerable environment which can effectively induce noise 
into our system's dynamics. For 
very small values of $\gamma_0$, we have shown that the GP computed is similar to the unitary GP, independently
of low or high frequency fluctuations.
We have also noted that the  difference between both phases increases for stronger 
values of $\gamma_0$, becoming important when there are low frequency 
longitudinal fluctuations in the environment. The difference among the phases are not considerable
if the fluctuations of low frequency are originated in a transversal noise ($\gamma_1$).
The correction to the GP is almost 
imperceptible to  the transversal fluctuations, at least in the weak coupling limit. 

It is important to recall that the results presented show 
that the system evolution in
the presence of an environment with high frequency fluctuations is very similar to the unitary evolution, since
the GP aquired is practically similar to the $\Phi_U$, for almost all values of $\gamma_0$. 
We believe that these results show a very similar scenario to that of the experimental situation reported
in \cite{scienceexp} where they have measured the Berry phase for a superconductiong qubit under 
high frequency fluctuations. In addition, we have checked that noise in the 
${\hat z}$-direction induces a bigger correction to the phase than the noise in the transversal components. 
This correction agrees with the previous analysis done on decoherence induced in
the qubit and with the experimental
setups reported of the observed geometric phase.  
Comparison between theory and experiment verifies our understanding of the physics 
underlying the system as a 
dissipative two-level device. 
The analysis of the dephasing time-scales may provide additional information
about the statistical properties of the 
noise. Berry's phase measurements provide an important constraint to take into 
account about noise models 
and their correction induced over the GP, at least, at the times in which the experiments can be performed. 
The comprehension of the
decoherence and dissipative processes should allow their further suppression in future qubits designs or 
experimental setups.


\acknowledgments

This work is supported by CONICET, UBA, and ANPCyT, Argentina.

\end{document}